\documentclass[sn-mathphys-num]{sn-jnl}

\usepackage{graphicx}%
\usepackage{multirow}%
\usepackage{amsmath,amssymb,amsfonts}%
\usepackage{amsthm}%
\usepackage{mathrsfs}%
\usepackage[title]{appendix}%
\usepackage{xcolor}%
\usepackage{textcomp}%
\usepackage{manyfoot}%
\usepackage{booktabs}%
\usepackage{algorithm}%
\usepackage{algorithmicx}%
\usepackage{algpseudocode}%
\usepackage{listings}%
\usepackage{braket}

\newcommand{\tr}{\mathrm{tr}\,}
\newcommand{\Imp}{\mathrm{Im}\,}

\newcommand{\Ad}{\mathrm{Ad}}
\newcommand{\Rep}{\mathrm{Re}}

\theoremstyle{thmstyleone}%
\theoremstyle{thmstyletwo}%

\theoremstyle{thmstylethree}%

\begin{document}

\title[The Jaynes-Cummings model in Phase Space Quantum Mechanics]{The Jaynes-Cummings model in Phase Space Quantum Mechanics}


\author[1]{\fnm{Mar} \sur{S\'anchez--C\'ordova}}\email{maria.cordova@uaslp.mx}
\equalcont{These authors contributed equally to this work.}

\author*[1]{\fnm{Jasel} \sur{ Berra--Montiel}}\email{jasel.berra@uaslp.mx}
\equalcont{These authors contributed equally to this work.}

\author[1]{\fnm{Alberto} \sur{Molgado}}\email{alberto.molgado@uaslp.mx}
\equalcont{These authors contributed equally to this work.}

\affil*[1]{\orgdiv{Facultad de Ciencias}, \orgname{Universidad Aut\'onoma de San Luis 
	Potos\'{\i}}, \orgaddress{\street{Campus Pedregal, Av. Parque Chapultepec 1610, Col. Privadas del Pedregal}, \city{San
	Luis Potos\'{\i}}, \postcode{78217}, \state{SLP}, \country{M\'exico}}}


\abstract{In this paper,  we address the phase space formulation of the Jaynes–Cummings model through the explicit construction of the full Wigner function for a hybrid bipartite quantum system composed of a two-level atom and a quantized coherent field. By employing the Stratonovich–Weyl correspondence and the coadjoint orbit method, we derive an informationally complete quasi-probability distribution that captures the full dynamics of light–matter interaction. This approach provides a detailed phase space perspective of fundamental quantum phenomena such as Rabi oscillations, atomic population inversion, and entanglement generation. We further measure the purity of the reduced quantized  field state by means of an appropriate Wigner function corresponding to the bosonic field part in order to investigate the entanglement dynamics of the system.}

\keywords{Phase space quantum mechanics,  Jaynes-Cummings model, Wigner function, light-matter interaction}



\maketitle
\section{Introduction}
In recent years, the phase space formulation of quantum mechanics has experienced a significant revival, particularly within the rapid evolving landscape of quantum technologies \cite{Overview, Advances}. This framework, inherently tied to the deformation quantization approach \cite{Curtright,Cosmas}, provides a powerful and intuitive quasi-probabilistic formulation for analyzing quantum systems, effectively connecting classical and quantum descriptions. Notably, phase space techniques have become central for the rigorous characterization of quantum states, especially through quantum state tomography, an essential tool for the validation of quantum devices and the certification of quantum information protocols \cite{purity,Hradil,James}. Beyond its applications in quantum information, this approach has found increasing relevance in the study of quantum fields \cite{Manko1,Manko2,Tomography}, relativistic quantum information \cite{RQI1,RQI2,RQI3}, and even in different approaches of quantum gravity \cite{LQG1,LQG2,LQG3}.
Within this context, the study of hybrid quantum systems, combining discrete and continuous  degrees of freedom, such as the interaction between a qubit and a bosonic field, have become a focal point of research. Their relevance stems from their central role in quantum communication and metrology, where the dynamics of coherence and entanglement are intricately shaped by light–matter interactions \cite{Cohen,Vahala}.

This manuscript focuses on the explicit construction of the full Wigner function for a canonical representative hybrid quantum system, the Jaynes–Cummings (JC) model, which describes the coherent interaction between a two-level atom and a single-mode quantized electromagnetic field \cite{Jaynes,Knight}. By employing the Stratonovich–Weyl formalism, we derive an informationally complete Wigner function that encapsulates the full dynamical behavior of the system, particularly for initial states composed of a two-level system (such as an atom or a qubit) and a quantized coherent field. Coherent states play a fundamental role in this context, as they represent the quantum states most closely resembling classical electromagnetic fields, both in structure and evolution \cite{Perelomov, Glauber}. This formulation enables a detailed phase space analysis of fundamental quantum optical phenomena such as Rabi oscillations, atomic population inversion, and the collapse and revival of coherence \cite{Meystre}. Furthermore, we explore the generation of quantum correlations and entanglement through the computation of reduced Wigner functions and purity measures.
Motivated by recent advances in the use of generalized Wigner functions to characterize non-classicality, particularly through the negativity volume as an entanglement witness for hybrid bipartite states \cite{Rundle, Arkhipov}. The results presented here not only offer a deeper understanding of the Jaynes–Cummings (JC) model from a phase space perspective but also a practical framework for capturing the entanglement dynamics. Notably, since the direct measurement of the Wigner function in phase space can be experimentally more accessible than the full tomographic reconstruction of the corresponding density matrix, our approach provides a convenient and efficient tool for identifying entanglement in hybrid systems.

The paper is organized as follows. In Section~\ref{sec:PSS}, we review the phase space quantization framework for hybrid bipartite systems. Section~\ref{sec:JCModel} introduces the full Wigner function for the Jaynes–Cummings model. This quasi-probability distribution enables a detailed characterization of the dynamics of the light–matter interaction, including the analysis of Rabi oscillations and atomic population inversion. Furthermore, by computing the purity from the reduced Wigner function, we examine the entanglement dynamics of the system. Concluding remarks and future perspectives are presented in Section~\ref{sec:Conclusions}.

\section{Phase space quantization and the Stratonovich–Weyl Map for hybrid bipartite qubit-bosonic states}
\label{sec:PSS}

The classical phase space for spin systems can be rigorously constructed using the coadjoint orbit method \cite{Kostant, Kirillov}, a geometric approach rooted in the representation theory for Lie groups. In the case of spin degrees of freedom, the relevant group is given by $SU(2)$, whose coadjoint orbits in the dual of its Lie algebra, $\mathfrak{su}(2)^{*}$, are diffeomorphic to two-dimensional spheres $S^{2}$. Each coadjoint orbit corresponds to a fixed value of the quadratic Casimir invariant which carries a natural symplectic structure known as the Kirillov–Kostant–Souriau (KKS) form \cite{Orbit}. These orbits provide a natural compact symplectic manifold structure for describing spin systems from the symmetry group itself and offers a geometrically consistent route to the phase space quantization via the Stratonovich–Weyl map, as detailed below.

\noindent Let $G=SU(2)$, whose Lie algebra $\mathfrak{g}=\mathfrak{su}(2)$ satisfies the commutation relations 
\begin{equation}
    \left[\sigma_{i},\sigma_{j}\right]=2i\epsilon_{ijk}\sigma_{k},
\end{equation}
for $i,j,k\in \{x,y,z\}$. These generators correspond to the angular momentum operators, and in the fundamental spin-1/2 representation they are realized by the Pauli matrices $\sigma_{i}$. The dual space $\mathfrak{su}(2)^{*}$ carries a natural Poisson structure, known as the Lie-Poisson structure,  which turns $\mathfrak{su}(2)^{*}$ into a Poisson manifold \cite{Vaisman}. The symplectic leaves of this Poisson manifold are exactly the coadjoint orbits of the Lie group $SU(2)$, each of which is a two-dimensional sphere corresponding to a fixed value of the Casimir invariant. The coadjoint orbit through a point $\xi\in\mathfrak{su}(2)^{*}$ is defined as the set 
\begin{equation}
    \mathcal{O}_{\xi}=\{\Ad_{g}^{*}\,\xi\;|\;g\in SU(2)\},
\end{equation}
where the map $\Ad_{g}^{*}:\mathfrak{g}^{*}\to\mathfrak{g}^{*}$ denotes the coadjoint action of the Lie group on the dual of its Lie algebra. The coadjoint orbit $\mathcal{O}_{\xi}$ is shown to be a smooth manifold diffeomorphic to a two-sphere $S^{2}$, where the symplectic KKS form on this orbit  can be expressed in local spherical coordinates $(\theta,\phi)$ as
\begin{equation}
    \omega_{\mathrm{symp}}=\frac{1}{2}\sin\theta d\theta\wedge d\phi.
\end{equation}

The quantization on coadjoint orbits proceeds by associating classical observables, i.e., smooth functions on the orbit, with quantum operators on a Hilbert space. This is accomplished by introducing a map known as the Stratonovich-Weyl (SW) correspondence which provides a consistent and covariant framework for such an association. In the case of spin-1/2 systems, the Hilbert space is given by $\mathbb{C}^{2}$ and, consequently, the space of quantum observables corresponds to the set of $2\times 2$ Hermitian matrices. The SW correspondence is implemented via a family of operator-valued distributions $\hat{\Delta}(\Omega)$, referred to as the Stratonovich–Weyl kernel, where $\Omega$ denotes any parametrization of the classical phase space, typically specified by coordinates on the two-sphere $S^{2}$. This kernel is required to satisfy the following physically motivated postulates \cite{Stratonovich}: 
\begin{enumerate}
	\item The mapping $W_{\hat{A}}(\Omega)=\tr\left[ \hat{A}\hat{\Delta}(\Omega)\right] $ defines a one-to-one linear map, where $W_{\hat{A}}(\Omega)$ represents the phase space function associated to the operator $\hat{A}$.
	\item The function $W_{\hat{A}}(\Omega)$ is real-valued, which implies that the operator kernel $\hat{\Delta}(\Omega)$ must be Hermitian.
	\item $W_{\hat{A}}(\Omega)$ satisfies the standarization condition, meaning that its integral over the entire phase space $\int_{\Omega}W_{\hat{A}}(\Omega)d\Omega=\tr \hat{A}$ exists and $\int_{\Omega}\hat{\Delta}(\Omega)d\Omega=\hat{1}$.
	\item Traciality, that is $\int_{\Omega}W_{\hat{A}}(\Omega)W_{\hat{B}}(\Omega)d\Omega=\tr\left[\hat{A}\hat{B} \right] $.
	\item Covariance, which ensures that if an operator $\hat{A}$ is invariant under global unitary operators, then its corresponding phase space function $W_{\hat{A}}(\Omega)$ also remains invariant. In other words, symmetry under conjugation by unitary operators is preserved under the Stratonovich–Weyl correspondence.
\end{enumerate}  

On the one hand, for the case of a two level spin-1/2 quantum system, such as a qubit represented over the Bloch sphere, the SW kernel reads \cite{Varilly, Tilma}, 
\begin{eqnarray}\label{Delta}
	\hat{\Delta}_q(\theta,\phi)&=&\frac{1}{2}\hat{U}(\phi,\theta,\Phi)\hat{\Pi}_q\hat{U}(\phi,\theta,\Phi)^{\dagger}, \nonumber\\
	&=&\frac{1}{2}\left(
	\begin{array}{cc}
		1+\sqrt{3}\cos{\theta} & \sqrt{3}e^{i\phi}\sin{\theta} \\
		\sqrt{3}e^{-i\phi}\sin{\theta} & 1-\sqrt{3}\cos{\theta} 
	\end{array}
	\right).		 
\end{eqnarray}
where $0\leq\theta\leq \pi$, $0\leq\phi\leq 2\pi$ and $0\leq\Phi\leq 4\pi$ are the conventional Euler angles \cite{Sakurai}.
In this expression $\hat{\Pi}_{q}$ denotes the parity operator, defined by
\begin{equation}\label{Parity}
	\hat{\Pi}_q=\hat{1}_{2}+\sqrt{3}\hat{\sigma}_z, 
\end{equation}
 where $\hat{1}_{2}$ is the identity matrix $2\times 2$ and $\hat{\sigma}_{z}$ represents the Pauli z operator. This particular form of the spin-parity operator arises from analyzing the $SU(N)$ coherent states on the complex projective space, as discussed in \cite{Perelomov,Nemoto}. The operator $\hat{U}(\phi,\theta,\Phi)$ in Eq.~\ref{Delta} plays an analogous  role to the displacement operator in continuous-variables systems. It corresponds to the general $SU(2)$ rotation operator, given by the Euler angle decomposition:  
\begin{equation}\label{SU2Rotation}
	\hat{U}(\phi,\theta,\Phi)=exp\left(-i\hat{\sigma}_z\frac{\phi}{2}\right)exp\left(-i\hat{\sigma}_y \frac{\theta}{2}\right)exp\left(-i\hat{\sigma}_z\frac{\Phi}{2}\right).
\end{equation}
It is noteworthy that the kernel $\hat{\Delta}_q(\theta,\phi)$ does not depend explicitly on the parameter $\Phi$. This independence reflects the gauge freedom in the Euler angle parametrization and allows us to restrict the associated phase space functions precisely to the coadjoint orbit corresponding to the two-sphere $S^{2}$.

On the other hand, for the case of continuous-variable systems, such as a bosonic field, the SW kernel operator $\hat{\Delta}(\Omega)$ is naturally expressed in terms of complex phase space coordinates $\Omega=\{\alpha,\alpha^{*}\}$. In this setting, the SW kernel takes the form
\begin{equation}
\label{eq:ker-field}
\hat{\Delta}_{f}(\alpha,\alpha^{*})=\frac{2}{\pi}\hat{D}(\alpha,\alpha^{*})\hat{\Pi}_{f}\hat{D}(\alpha,\alpha^{*})^{\dagger},
\end{equation}
where $\hat{D}(\alpha,\alpha^{*})$ corresponds to the displacement operator that generates coherent states, defined as
\begin{equation}
\hat{D}(\alpha,\alpha^{*})=\exp(\alpha\hat{a}^{\dagger}-\alpha^{*}\hat{a}),
\end{equation}
with $\hat{a}^{\dagger}$ and $\hat{a}$ denoting the bosonic creation and annihilation operators, respectively. The operator $\hat{\Pi}_{f}$ appearing in this expression is the bosonic parity operator, given by
\begin{equation}
\hat{\Pi}_{f}=\exp(i\pi\hat{a}^{\dagger}\hat{a})=\frac{1}{\pi}\int\ket{\alpha}\bra{-\alpha}d^{2}\alpha,
\end{equation}
where $\ket{\alpha}=\hat{D}(\alpha,\alpha^{*})\ket{0}$ denotes a coherent state, and the measure is defined by $d^{2}\alpha=d\Rep(\alpha)d\Imp(\alpha)$. 

With kernels~(\ref{Delta}) and~(\ref{eq:ker-field}) at hand, we can construct the SW correspondence for hybrid quantum systems  \cite{Rundle, Arkhipov}. In particular for the case consisting of a bosonic field interacting with a two-level quantum system (qubit), the SW kernel reads
\begin{equation}
    \hat{\Delta}(\Omega)=\hat{\Delta}_q(\theta,\phi)\otimes\hat{\Delta}_f(\alpha,\alpha^{*}).
\end{equation}
This expression implies that the Wigner function corresponding to a hybrid quantum state $\hat{\rho}$ is written as
\begin{equation}\label{SWQ}
	W_{\hat{\rho}}(\theta,\phi,\alpha,\alpha^{*})=\tr\left[\hat{\rho}\,\hat{\Delta}_q(\theta,\phi)\otimes\\hat{\Delta}_f(\alpha,\alpha^{*})\right].
\end{equation} 
The standarization condition (iii) is ensured by integrating over the full hybrid phase space using the appropriate measure, which consists of the symplectic volume on the two-sphere $S^{2}$ for the qubit, and the measure on the coherent state space for the field $d^{2}\alpha$, thus obtaining
\begin{equation}
    d\Omega=\frac{1}{2\pi}\sin\theta\,d\theta d\phi\,d^{2}\alpha,
\end{equation}
with $0\leq\theta\leq \pi$, $0\leq\phi\leq 2\pi$, and $\Rep(\alpha),~\Imp( \alpha) \in \mathbb{R}$. 
The reduced density matrices for the qubit and bosonic field subsystems are obtained by performing a partial trace over the corresponding complementary degrees of freedom. Specifically, the reduced state of the qubit is given by $\hat{\rho}_{q}=\tr_{f}[\hat{\rho}]$, while the reduced state of the field is $\hat{\rho}_{f}=\tr_{q}[\hat{\rho}]$. The corresponding reduced Wigner functions are hence obtained by integrating out the subsystem in the full Wigner function. For the qubit subsystem, the reduced Wigner function reads
\begin{equation}
    W_{\hat{\rho}_{q}}(\theta,\phi)=\tr\left[\hat{\rho}_{q}\hat{\Delta}_{q}(\theta,\phi)\right]=\int W_{\hat{\rho}}(\theta,\phi,\alpha,,\alpha^{*})d^{2}\alpha,
\end{equation}
whereas for the bosonic field subsystem, the reduced Wigner function is given by:
\begin{equation}\label{reducedf}
    W_{\hat{\rho}_{f}}(\alpha,\alpha^{*})=\tr\left[\hat{\rho}_{f}\hat{\Delta}_{f}(\alpha,\alpha^{*})\right]=\frac{1}{2\pi}\int W_{\hat{\rho}}(\theta,\phi,\alpha,,\alpha^{*})\sin\theta d\theta d\phi.
\end{equation}
These expressions reflect the consistency of the Wigner formalism with the structure of reduced states obtained via the partial trace.

\section{The Wigner function of the Jaynes-Cummings model, Rabi oscillations and entanglement dyanmics}  
\label{sec:JCModel}

The Jaynes-Cummings (JC) model constitutes a fundamental theoretical framework in quantum optics that describes the coherent interaction between a two-level quantum system, typically, a qubit or an atom and a single mode of the quantized electromagnetic field. The JC model was originally introduced as a simplification of quantum electrodynamics, the JC model captures essential features of light–matter interaction under the rotating wave approximation \cite{Jaynes, Knight}. The JC Hamiltonian can be written as follows, setting $\hbar=1$ for simplicity,
\begin{equation}
\label{JC-Ham}
	\hat{H}=\omega\hat{1}_{2}\otimes\hat{a}^{\dagger}\hat{a}+\frac{\Omega}{2}\hat{\sigma}_3\otimes\hat{1}+g(\hat{\sigma}_{-}\otimes\hat{a}+\hat{\sigma}_+\otimes\hat{a}^\dagger).
\end{equation}
\noindent Here, $\omega$ denotes the frequency of the bosonic field, $\Omega$ represents the energy difference of the two-level quantum system and $g$ is the coupling strength between the field and the qubit. The operator $\hat{1}_{2}$ is the $2\times 2$ identity matrix acting on the qubit Hilbert space, while $\hat{1}$ denotes the identity operator on the field mode. And the atomic operators $\hat{\sigma_{\pm}}$ in the last term are customarily defined as $(\hat{\sigma}_{x}\pm i \hat{\sigma}_{y})/2$. 
A standard analysis of the Jaynes–Cummings model involves applying the JC Hamiltonian~(\ref{JC-Ham}) to an initial coherent state, which is a minimum-uncertainty Gaussian state in phase space \cite{Glauber}. A key feature of this model is that the interaction drives the field into a Schr\"odinger cat state. This characteristic makes the JC setup particularly important for applications in continuous-variable quantum information, thus providing a fertile ground for testing concepts in entanglement dynamics and decoherence.

To determine the time evolution of the system, we need to solve the Schr\"odinger equation 
\begin{equation}\label{Schrodinger}
	i\frac{d\hat{U}(t)}{dt}=\hat{H}\hat{U}(t)=(\hat{H}_{0}+\hat{V})\hat{U}(t),
\end{equation} 
where $\hat{U}(t)$ is the unitary evolution operator of the composite system, $\hat{H}_{0}$ denotes the free Hamiltonian describing the qubit and the bosonic field independently, and $\hat{V}$ represents the interaction term, corresponding to the first two terms and the last term in the JC Hamiltonian~(\ref{JC-Ham}), respectively. Under the resonance condition, $\Omega=\omega$, this equation can be explicitly solved by using the method of constant variation \cite{Meystre,Fujii,Moya}, yielding the solution
\begin{equation}\label{evol}
	\hat{U}(t)=(e^{-it\frac{\omega}{2}\hat{\sigma}_3} \otimes e^{-itw\hat{N}})(e^{-itg(\hat{\sigma}_{-}\otimes\hat{a}+ \hat{\sigma}_{+}\otimes\hat{a}^\dagger)}).
\end{equation}
In this resonant regime, energy can be exchanged most efficiently between the atom and the field, resulting in coherent Rabi oscillations with a maximum amplitude \cite{Meystre}. 
The explicit matrix representation of $\hat{U}(t)$ in the qubit basis is given by
\begin{equation}
	\hat{U}(t)=
	\left(
	\begin{array}{cc}
		e^{-iwt(\hat{N}+\frac{1}{2})}\cos(tg\sqrt{\hat{N}+1}) & -ie^{-iwt(\hat{N}+\frac{1}{2})} \frac{\sin(tg\sqrt{\hat{N}+1})}{\sqrt{\hat{N}+1}}\hat{a} \\
		-ie^{-iwt(\hat{N}-\frac{1}{2})}\frac{\sin(tg\sqrt{\hat{N}})}{\sqrt{\hat{N}}}\hat{a}^\dagger & e^{-iwt(\hat{N}-\frac{1}{2})}\cos(tg\sqrt{\hat{N}})
	\end{array}
	\right),	 
\end{equation}
where $\hat{N}=\hat{a}^{\dagger}\hat{a}$ stands for the number operator of the bosonic field, such that $\hat{N}\ket{n}=n\ket{n}$. To investigate the dynamics governed by the JC model, we consider the initial state of the composite system given by
\begin{equation}\label{initial}
	\ket{\psi_{0}}=\left(C_e\ket{e}+ C_g\ket{g}\right)\otimes\ket{\alpha},
\end{equation}
where $C_{e}$ and $C_{g}$ are complex coefficients satisfying $|C_{e}|^{2}+|C_{g}|^{2}=1$, $\ket{e}$ and $\ket{g}$ denote the excited and ground states of the two-level system, respectively, while $\ket{\alpha}$ is a coherent state of the quantized field mode. This choice is physically well-motivated, since coherent states represent the most classical-like states of the electromagnetic field and can be readily prepared in optical cavities, while the atom (or qubit) is initialized in a general pure superposition, as it is typically done via external control fields \cite{JC, Cavity}.  Moreover, this separable initial condition  avors an unambiguous analysis of the entanglement and energy exchange dynamics that are entirely induced by the atom–field interaction \cite{EntanglementJC}. After applying the evolution operator (\ref{evol}) to the initial state (\ref{initial}), the resulting state of the system reads
\begin{eqnarray}
	\mkern-60mu\ket{\psi(t)}&=&\sum_{n=0}^{\infty}\left([C_{n}C_{e} e^{-it E_{n}}\cos{(tg\sqrt{n+1})}-iC_{n+1}C_{g}e^{-it E_{n}}\sin{(tg\sqrt{n+1})}]\ket{e,n}\right.
	\nonumber\\ 
    && + \left.[C_{n}C_{g} e^{-it E_{n-1}}\cos{(tg\sqrt{n})}-iC_{n-1}C_{e} e^{it E_{n-1}}\sin{(tg\sqrt{n})}]\ket{g,n}\right).
\end{eqnarray}
where $E_{n}=\omega \left(n+1/2\right)$ and $E_{n-1}=\omega \left(n-1/2\right)$. The coefficients $C_{n}$ corresponds to the amplitudes of the coherent state $\ket{\alpha}$ when expressed in the Fock basis as
\begin{equation}\label{coeff}
    \ket{\alpha}=e^{-|\alpha|^{2}/2}\sum_{n=0}^{\infty}\frac{\alpha^{n}}{\sqrt{n!}}\ket{n}\equiv \sum_{n=0}^{\infty}C_{n}\ket{n},
\end{equation}
and satisfy the normalization condition $\sum_{n}^{\infty}|C_{n}|^{2}=1$. In the context of the JC model, it is common to consider the initial state of the atom to be fully excited, corresponding to the choice $C_{e}=1$ and $C_{g}=0$. This selection simplifies the analysis while retaining the essential features of the atom–field interaction, particularly in the study of Rabi oscillations. As we will see below, these conditions enable a clear observation of excitation exchange between the atom and the quantized field modes. Moreover, this fully excited configuration illustrates the phenomenon of collapse and revival in the atomic population inversion, which arises from the discrete structure of the photon number distribution inherent to the field's quantum state \cite{Knight, Meystre}. For this specific configuration, the density operator is given by 
\begin{eqnarray}\label{density}
\hat{\rho}&=&\sum_{n,m=0}^{\infty} C_n C_m^* e^{-i t \omega (E_n - E_m)} \Big[  
\cos(tg\sqrt{n+1}) \cos(tg\sqrt{m+1}) \ket{e,n}\bra{e,m} \nonumber \\
&&+ i \cos(tg\sqrt{n+1}) \sin(tg\sqrt{m+1}) \ket{e,n}\bra{g,m+1} \nonumber \\
&&- i \sin(tg\sqrt{n+1}) \cos(tg\sqrt{m+1}) \ket{g,n+1}\bra{e,m} \nonumber \\
&&+ \sin(tg\sqrt{n+1}) \sin(tg\sqrt{m+1}) \ket{g,n+1}\bra{g,m+1} \Big].
\end{eqnarray}
\noindent Once the density operator is obtained from the time evolution governed by the JC Hamiltonian, the complete Wigner function for the hybrid bipartite quantum system can be constructed using the Stratonovich-Weyl map (\ref{SWQ}). Since the density operator (\ref{density}) is defined on the tensor product Hilbert spaces $\mathcal{H}_{q}\otimes \mathcal{H}_{f}$, we can make use of the property that the trace of a tensor product factorizes as $\tr(\hat{A}\otimes \hat{B})=\tr(\hat{A}) \tr(\hat{B})$, where $\hat{A}$ and $\hat{B}$ act on independent Hilbert spaces. Then, the Wigner function associated to density operator  reads
\begin{eqnarray}\label{Wignerb}
W_{\hat{\rho}}(\theta, \phi, \beta,\beta^{*})&=&\sum_{n,m=0}^{\infty}\Big[\frac{1}{2}(1+\sqrt{3}\cos\theta) C_n{C_m}^{*} {e^{-it(E_n-E_m)}}\cos(tg\sqrt{n+1})\cos(tg\sqrt{m+1})\nonumber\\
&&+i\frac{\sqrt{3}}{2}e^{i\phi}\sin\theta C_n C_{m-1}^{*}{e^{-it(E_n-E_{m-1})}}\cos(tg\sqrt{n+1})\sin(tg\sqrt{m})\nonumber\\
&&-i\frac{\sqrt{3}}{2}e^{-i\phi}\sin\theta C_{n-1}C_m^{*} {e^{-it(E_{n-1}-E_m)}}\sin(tg\sqrt{n})\cos(tg\sqrt{m+1})\nonumber\\
&&+\frac{1}{2}(1-\sqrt{3}\cos\theta) C_{n-1}C_{m-1}^{*} {e^{-it(E_{n-1}-E_{m-1})}}\sin(tg\sqrt{n})\sin(tg\sqrt{m})\Big]\nonumber\\
&&\times\ \frac{2}{\pi}\frac{e^{-2|\beta|^{2}}}{\sqrt{n!m!}}L_{n,m}(2\beta,2\beta^{*}),
\end{eqnarray}		
where $L_{n,m}(\beta,\beta^{*})$ correspond to the Laguerre 2D polynomials \cite{Wunsche}, defined by the operational representation
\begin{eqnarray}
    L_{n,m}(\beta,\beta^{*})&=&\exp\left(-\frac{\partial^{2}}{\partial\beta\partial\beta^*}\right)\beta^{n}\beta^{*m}, \nonumber \\
    &=&\sum_{j=0}^{n,m}\frac{n!m!}{j!(n-j)!(m-j)!}(-1)^{j}\beta^{n-j}\beta^{*m-j}.
\end{eqnarray}

\noindent By employing the identity \cite{Laguerre},
\begin{equation}
    \frac{1}{\pi}\int e^{r\frac{|\beta|^{2}}{2}}e^{-\frac{|\beta|^{2}}{2}}L_{n,m}(\beta,\beta^{*})d^{2}\beta=\frac{2\pi}{1-r} \sqrt{m!\,n!}\left(\frac{1+r}{1-r}\right)^{n}\delta_{n,m},
\end{equation}
and the orthogonality relation,
\begin{equation}\label{ortho}
    \frac{1}{\pi}\int e^{-|\beta|^{2}}L_{m,l}(\beta,\beta^{*})L_{k,n}(\beta,\beta^{*})d^{2}\beta=m!\,l!\,\delta_{m,n}\delta_{l,k}
\end{equation}
one can verify that the Wigner function of the hybrid bipartite quantum system, defined in Eqn.~(\ref{Wignerb}), is both real and properly normalized. Specifically, it satisfies the normalization condition
\begin{equation}
    \frac{1}{2\pi}\int W_{\hat{\rho}}(\theta, \phi, \beta,\beta^{*})\sin\theta d\theta d\phi d^{2}\beta=1.
\end{equation}
 These properties ensure that the Wigner function provides a valid quasi-probability distribution over the combined phase space of the spin and bosonic subsystems.

Let us now compute the Wigner function corresponding to the one-mode oscillation in the Jaynes–Cummings model, where the initial state is $\ket{e,r}$ and the system evolves coherently into $\ket{g,r+1}$. This particular case encapsulates the fundamental mechanism of excitation exchange between a two-level system and a single quantized bosonic mode. To isolate this specific case, we assume that the coefficients satisfy $C_{n}=\delta_{n,r}$ and $C_{m}=\delta_{m,r}$,  corresponding to an initial field state in the Fock state $\ket{r}$. Under this assumption, the general expression for the Wigner function given in Eqn.~(\ref{Wignerb}) simplifies to:
\begin{eqnarray}\label{Wignern}
\mkern-30mu W_{\hat{\rho}_{r}} (\theta, \phi, \beta,\beta^{*})&=&
\Big[-\frac{\sqrt{3}}{2}\sin{\theta} \sin{\phi}\sin{(2tg\sqrt{r+1})}+\frac{1}{2}+\nonumber\\
&& \frac{\sqrt{3}}{2}\cos{\theta}\cos{(2tg\sqrt{r+1})}\Big] \Big(\frac{2}{\pi}e^{-2|\beta|^{2}}(-1)^{r}L_{r}(4|\beta|^{2})\Big),
\end{eqnarray}
where $L_{r}(\beta)$ denotes the standard Laguerre polynomials of degree $r$, introduced by the identity $L_{r,r}(\beta,\beta^{*})=(-1)^{r}r! L_{r}(|\beta|^{2})$. From this expression, we observe that the last term corresponds precisely to the Wigner quasi-probability distribution associated with the one-mode Fock state $\ket{r}$ \cite{Knight}.  

Rabi oscillations describe the coherent and periodic exchange of excitation between a two-level atom and a quantized mode of the bosonic field, manifesting as oscillations in the atomic population inversion, which quantifies the difference in probability between the atom being in the excited or ground state \cite{Meystre,Merlin}. Within the Wigner formalism, these dynamics can be visualized and analyzed in phase space, offering a quasi-probabilistic representation of the quantum state evolution. The Wigner function allows for a clear depiction of interference patterns, non-classical features, and phase correlations between the atom and field.  For initial states consisting of a coherent field coupled to an excited atom, the resulting Rabi oscillations exhibit the characteristic collapse and revival phenomena in the atomic inversion. Within the phase space formalism, these dynamical features can be extracted directly from the quasi-probability distribution given in (\ref{Wignern}), as detailed below. To determine the probability that the quantum system remains in the excited state, we compute the Stratonovich-Weyl transform of the projection operator $\hat{P}_{\ket{e,r}}=\ket{e,r}\bra{e,r}$. Using Eqn. (\ref{SWQ}) we obtain
\begin{eqnarray}
    W_{\hat{P}_{\ket{e,r}}}(\theta,\phi,\beta,\beta^{*})&=&\tr\left[\hat{P}_{\ket{e,r}}\,\hat{\Delta}_q(\theta,\phi)\otimes\hat{\Delta}_f(\beta) \right], \nonumber \\
    &=& \frac{1}{\pi}(1+\sqrt{3}\cos\theta)e^{-2|\beta|^{2}}(-1)^{r}L_{r}(4|\beta|^{2}).
\end{eqnarray}
Then, the probability that the system remains in the excited state at time $t$ is given by integration of the SW symbol of the projection operator and the Wigner function of the total state,
\begin{eqnarray}
    P_{e}(t)&=&\frac{1}{2\pi}\int W_{\hat{P}_{\ket{e,r}}}(\theta,\phi,\beta,\beta^{*})W_{\hat{\rho}_{r}} (\theta, \phi, \beta,\beta^{*})\sin\theta d\theta d\phi d^{2}\beta, \nonumber \\
    &=& \cos^{2}(tg\sqrt{r+1}).
\end{eqnarray}
Similarly, the probability that the system is found in the ground state is given by
\begin{eqnarray}
P_{g}(t)&=&\frac{1}{2\pi}\int W_{\hat{P}_{\ket{g,r+1}}}(\theta,\phi,\beta,\beta^{*})W_{\hat{\rho}_{r}} (\theta, \phi, \beta,\beta^{*})\sin\theta d\theta d\phi d^{2}\beta, \nonumber \\
    &=& \sin^{2}{(tg\sqrt{r+1})}.
\end{eqnarray}    
The atomic inversion defined as the difference between the excited and ground state probabilities thus reads
\begin{equation}
\mkern-70mu Z(t)=P_e{(t)}-P_g{(t)}=\cos^{2}{(tg\sqrt{r+1})}-\sin^{2}{(tg\sqrt{r+1})} 
=\cos(2tg\sqrt{r+1}).
\end{equation}
These are the Rabi oscillations, which characterize the coherent exchange of energy between the two-level atom and the quantized field mode. In the one-mode scenario, this simple configuration captures the essence of light–matter interaction under the Jaynes–Cummings model and serves as an example for understanding more complex dynamical features such as quantum-classical transition, decoherence and quantum noise \cite{Coherence, Light}. Moreover, the explicit sinusoidal form of the atomic inversion function $Z(t)$  highlights the dependence of the oscillation frequency on the number state of the field, thereby emphasizing the quantum nature of the bosonic field and the nonlinear atom–field coupling characteristic of this light-matter interaction model.

We now extend the analysis to the case of a coherent multi-mode field interacting with a two-level atom. By employing the full bipartite hybrid Wigner function, introduced in Eqn.~(\ref{Wignerb}), the probability that the system remains in the excited state can be expressed as
\begin{eqnarray}
 P_{e}(t)&=&\frac{1}{2\pi}\int W_{\hat{P}_{\ket{e}}}(\theta,\phi,\beta,\beta^{*})W_{\hat{\rho}} (\theta, \phi, \beta,\beta^{*})\sin\theta d\theta d\phi d^{2}\beta, \nonumber \\
    &=& \sum_{n=0}^{\infty} |C_n|^{2} \cos^{2}(tg\sqrt{n+1}),
\end{eqnarray}
where the coefficients $C_{n}$ represents the amplitudes of the initial coherent state, as defined in (\ref{coeff}) and $ W_{\hat{P}_{\ket{e}}}(\theta,\phi,\beta,\beta^{*})$ denotes the Stratonovich-Weyl transform of the projection operator $\hat{P}_{\ket{e}}=\sum_{m,n}\ket{e,m}\bra{e,n}$.
Analogously, the probability that the system is found in the ground state is given by
\begin{eqnarray}
 P_{g}(t)&=&\frac{1}{2\pi}\int W_{\hat{P}_{\ket{g}}}(\theta,\phi,\beta,\beta^{*})W_{\hat{\rho}} (\theta, \phi, \beta,\beta^{*})\sin\theta d\theta d\phi d^{2}\beta, \nonumber \\
    &=& \sum_{n=0}^{\infty} |C_n|^{2} \sin^{2}(tg\sqrt{n+1}),
\end{eqnarray}
In this case, the atomic inversion reads
\begin{equation}
Z(t)=P_e{(t)}-P_g{(t)}=\sum_{n=0}^{\infty} |C_n|^{2}\cos(2tg\sqrt{n+1}).
\end{equation}
This expression represents a weighted sum of sinusoidal oscillations with incommensurate frequencies, i.e., frequencies that are not integer multiples of a common base frequency. The resulting dynamics exhibits the characteristic signatures of the collapse and revival phenomena, where the atom–field interaction initially appears to loose coherence due to the dephasing of contributions from different field-number states, followed by a partial rephasing of the mode contributions, leading to the well-known revival of coherent oscillations at times
\begin{equation}
\label{RevTime}
    T_{\mathrm{rev}}\approx \frac{2\pi k}{g}\sqrt{\braket{\hat{N}}},
\end{equation}
where $k\in \{0,1,2,\ldots\}$ and $\braket{\hat{N}}=|\alpha|^{2}$ is the average number of excitations in the field. For further details and a detailed discussion on this topic, see, for instance \cite{ Knight, Meystre}.

We conclude this section with a discussion on the entanglement dynamics, based on the hybrid bipartite Wigner function computed for the Jaynes–Cummings model. From the total Wigner function (\ref{Wignerb}), we observe that, due to the presence of the double summation, the system is generally not separable at arbitrary times.  In particular, the Wigner function cannot be factorized, and therefore 
\begin{equation}
W_{\hat{\rho}}(\theta, \phi, \beta,\beta^{*})\neq W_{\hat{\rho}_{q}}(\theta,\phi)W_{\hat{\rho}_{f}}(\beta,\beta^{*}).
\end{equation}
The appearance of cross terms in the summation reflects the existence of atom–field correlations throughout the time evolution. This non-factorizable structure of the Wigner function indicates the presence of quantum correlations, i.e., entanglement between the qubit and field subsystems. To quantify the entanglement in the system, we analyze the purity of the reduced state of the field \cite{purity, Nielsen}. A mixed reduced state implies that the total qubit–field system is entangled. As a first step, we consider the reduced Wigner function of the bosonic field, defined in Eq. (\ref{reducedf}), given by
\begin{eqnarray}
     W_{\hat{\rho}_{f}}(\beta,\beta^{*})&=&\frac{1}{2\pi}\int W_{\hat{\rho}}(\theta,\phi,\beta,\beta^{*})\sin\theta d\theta d\phi, \nonumber \\
     &=& \sum_{n,m=0}^{\infty}\Big[ C_n{C_m}^{*} {e^{-it(E_n-E_m)}}\cos(tg\sqrt{n+1})\cos(tg\sqrt{m+1})\nonumber\\
&&+C_{n-1}C_{m-1}^{*} {e^{-it(E_{n-1}-E_{m-1})}}\sin(tg\sqrt{n})\sin(tg\sqrt{m})\Big]\nonumber\\
&&\times\ \frac{2}{\pi}\frac{e^{-2|\beta|^{2}}}{\sqrt{n!m!}}L_{n,m}(2\beta,2\beta^{*}).
\end{eqnarray}
Within the phase space formalism, the purity of the reduced field state is defined as \cite{Gardiner},
\begin{equation}
    \xi(t)=\pi\int |W_{\hat{\rho}_{f}}(\beta,\beta^{*})|^{2}d^{2}\beta. 
\end{equation}
A value of $\xi=1$, indicates that the field is in a pure state, implying that the total qubit–field system is unentangled. Conversely, if $\xi<1$, the reduced field state is mixed, which signals the presence of entanglement in the total state. By making use of the orthogonality relation given in (\ref{ortho}), the purity of the reduced Wigner function can be expressed as 
\begin{eqnarray} \label{purity}
 \xi(t)&=& \frac{1}{2}+\frac{1}{2}\sum_{m,n}^{\infty}|C_{n}|^{2}|C_{m}|^{2}\cos(2gt(\sqrt{n+1}-\sqrt{m+1}))\nonumber \\
 && +\sum_{n,m}^{\infty}2\Rep \Big[C_{n}C_{n-1}^{*}C^{*}_{m}C_{m-1}\cos(gt\sqrt{n+1})\cos(gt\sqrt{m+1})\nonumber \\&&\times\sin(gt\sqrt{n}) \sin(gt\sqrt{m})\Big].
\end{eqnarray}
This expression coincides with the purity obtained from the reduced density matrix of the field, $\tr(\hat{\rho}_{f}^{2})$,
and quantifies the degree of mixedness of the field due to its entanglement with the qubit. Moreover, the contribution from the real part reflects quantum interference effects arising from the off-diagonal coherences between distinct number states. These interference terms are sensitive to the relative phases induced by the Jaynes–Cummings evolution. As time progresses, the contributions from these terms tend to average out due to dephasing, reflecting the loss of coherence and the transition of the field to an effectively mixed state.
It can be observed that at $t=0$, the system begins in an unentangled product state, as $\xi(0)=1$. However, for $t>0$, correlations and quantum interference arise between the qubit and field subsystems, signaling the development of entanglement. As each Fock component evolves with a distinct Rabi frequency, the coherence between the qubit and the field undergoes dephasing. This dephasing is reflected in the temporal behavior of the purity $\xi(t)$, which decreases as the reduced field state becomes mixed. At later times, specifically at the revival times $T_{rev}$~(\ref{RevTime}), the accumulated phases partially re-align, resulting in constructive interference and the partial restoration of coherence \cite{Knight}. This process leads to the revival of population inversion and, correspondingly, the revival of entanglement. During these oscillations, the system periodically approaches an almost separable state before becoming entangled again, exhibiting the characteristic collapse and revival dynamics of the Jaynes–Cummings model \cite{Meystre, JC}. In the long time limit, i.e., as $t\to\infty$, the interferece term in Eqn.~(\ref{purity}) average out to zero. Meanwhile, considering that the coefficients in Eqn.~(\ref{coeff})  correspond to a coherent state, the purity of the Wigner function asymptotically approaches to
\begin{equation}
    \lim_{t\to\infty}\xi(t)=\frac{1}{2}+\frac{1}{2}e^{-2|\alpha|^{2}}I_{0}(2|\alpha|^{2}),
\end{equation}
where $I_{0}(|\alpha|^{2})$ denotes the modified Bessel function of the first kind \cite{Bessel}. This implies that, although the quantum system begins in a pure state, the purity decays toward $1/2$ as $|\alpha|$ increases, approaching a maximally mixed state. This behavior reflects the growing entanglement and dephasing characteristic of the Jaynes–Cummings dynamics.
This result is consistent with the behavior of the entanglement entropy obtained via the density matrix formalism \cite{Phoenix,RundleE}, thereby confirming the effectiveness of the phase space description in capturing the entanglement dynamics.

\section{Conclusions}\label{sec:Conclusions}

In this paper, we have studied the phase space formulation of the Jaynes–Cummings model by constructing the full Wigner function for a hybrid quantum system composed of a two-level atom and a coherent quantized field. By using the Stratonovich–Weyl correspondence and the coadjoint orbit method, we derived an informationally complete quasi-probability distribution that captures the full dynamics of the light–matter interaction. This formulation provides a phase space perspective on relevant quantum phenomena such as Rabi oscillations, atomic population inversion, and entanglement generation. One of the main outcomes of this study is the observation that the Wigner function provides an effective and experimentally feasible method for tracking the entanglement dynamics between subsystems. In particular, we showed that the purity of the reduced Wigner function serves as a robust indicator of quantum correlations. This approach circumvents the need for a full state tomography, offering a practical advantage for experimental platforms such as cavity QED, trapped ion systems, and superconducting qubits. Furthermore, the use of coherent states as initial conditions connects naturally to physically realizable settings in quantum optics and quantum computing. We expect that the results established in this work, will pave the way for an explicit construction of the star product for hybrid bipartite systems, thereby enabling, in principle, the full machinery of the deformation quantization program applied to hybrid quantum dynamics. In particular, it would allow the incorporation of geometric techniques, such as symplectic connections, curvature corrections, and star-product operations within the hybrid phase space framework. This extension would not only deepen the theoretical understanding of light–matter interactions but also open new avenues for formulating advanced quantization schemes in complex quantum systems. Potential applications include the analysis of entanglement dynamics on phase space, the implementation of semiclassical approximations, and the investigation of non-classicality on generalized hybrid quantum systems.

\section*{Acknowledgments}
 The authors would like to acknowledge support from SNII CONAHCYT-Mexico.  MSC acknowledge financial support from SECIHTI under project Estancias Posdoctorales por México 2023(1). JBM acknowledge financial support from Marcos Moshinsky foundation.  AM acknowledges financial support from COPOCYT under project 2467 HCDC/2024/SE-02/16 (Convocatoria 2024-03, Fideicomiso 23871).

\section*{Data Availability Statement}
The authors declare that no data were generated or analyzed in the course of this research.

\bibliographystyle{unsrt}

\end{document}